\documentclass{iau-JDSS}
\usepackage{graphicx}
\title[Current Solar Minimum] 
{Peculiar Current Solar-Minimum \\ Structure of the Heliosphere}

\author[P.K. Manoharan]{P.K. Manoharan}

\affiliation{Radio Astronomy Centre, National Centre for Radio Astrophysics,\\
Tata Institute of Fundamental Research, Udhagamandalam (Ooty), 643001, India
\break email: mano@ncra.tifr.res.in}

\pubyear{2009}
\volume{Volume 15}  %% insert here IAU Highlights of Astronomy Volume Number
\pagerange{1--7}
\date{September 20, 2009 and in revised form September 21, 2009}
\jname{Highlights of Astronomy, Volume 15}
\editors{Ian F Corbett, ed.}

\newcommand{\kmps}{kms$^{-1}$}
\begin{document}

\maketitle

\begin{abstract}
In this paper, I review the results of 3-D evolution of 
the inner heliosphere over the solar cycle \#23, based on observations 
of interplanetary scintillation (IPS) made at 327 MHz using the Ooty 
Radio Telescope. The large-scale features of solar wind speed and 
density turbulence of the current minimum are remarkably different 
from that of the previous cycle. The results on the solar wind density 
turbulence show that (1) the current solar minimum is experiencing a low 
level of coronal density turbulence, to a present value of $\sim$50\% 
lower than the previous similar phase, and (2) the scattering diameter 
of the corona has decreased steadily after the year 2003. The results on 
solar wind speed are consistent with the magnetic-field strength at the 
poles and the warping of heliospheric current sheet.

\keywords{turbulence, scattering, Sun: corona, Sun: magnetic fields, 
Sun: coronal mass ejections (CMEs), solar wind, solar-terrestrial 
relations}
\end{abstract}


\firstsection % if your document starts with a section,

\section{Interplanetary Scintillation}

In this study, a large amount of interplanetary scintillation (IPS)
data obtained from the Ooty Radio Telescope (ORT), operating at 327 
MHz (Swarup et al. 1971), has been employed to study the 3-D evolution 
of the heliosphere over the period 1989--2009. The IPS observations
made with the ORT can provide the velocity of the solar wind and the 
scintillation index ({\it m}) in the heliocentric distance range 
of $R\sim$10--250 solar radii ($R_\odot$) and at all heliographic 
latitudes. The value of {\it m} is a measure of electron-density 
turbulence in the solar wind ($m^2 \sim \int\delta N^{2}_{e}(z)~dz$), 
along the line of sight ({\it z}) to the radio source (e.g.,  
Manoharan et al. 2000). The normalized scintillation index, 
$g = m(R)/{\rm<}m(R){\rm>}$ (i.e., observed index normalized by 
its long-term average), enables the comparison of levels of 
density turbulence obtained from different sources.
However, the value of {\it g} is linearly 
related to $\delta N_e$ only in the weak-scattering region at
distances $>$40 $R_\odot$. For example, an {\it m-R} profile attains 
the peak value at the strong-to-weak scattering transition point, 
which typically occurs $\sim$40 $R_\odot$ for IPS at 327 MHz (e.g., 
Manoharan 1993; 2006). In this study, the solar wind velocity and 
turbulence images have been exclusively obtained from weak-scattering 
data.  However, the contour of constant level of turbulence in a year
at different latitudes has been determined using peaks of several
{\it m-R} profiles. 

%\firstsection 
\section{Solar Cycle \#23: Three-Dimensional Solar Wind}
Figure 1a shows the latitudinal distributions of solar wind speed
and density turbulence ({\it g}) observed at Ooty over the solar 
cycle \#23. These plots are similar to the well-known `{\it 
butterfly diagram}' of photospheric magnetic field intensity.
They have been made by tracing backward/forward from the measurement 
location onto a sphere of radius $\sim$100 R$_\odot$, which 
approximately corresponds to the mid range of distances covered in
the observations utilized to generate the plots.

% small disatnces extrapolation may induce significant error on the
% distribution

\begin{figure}
\begin{center}
\includegraphics[width=6.7cm]{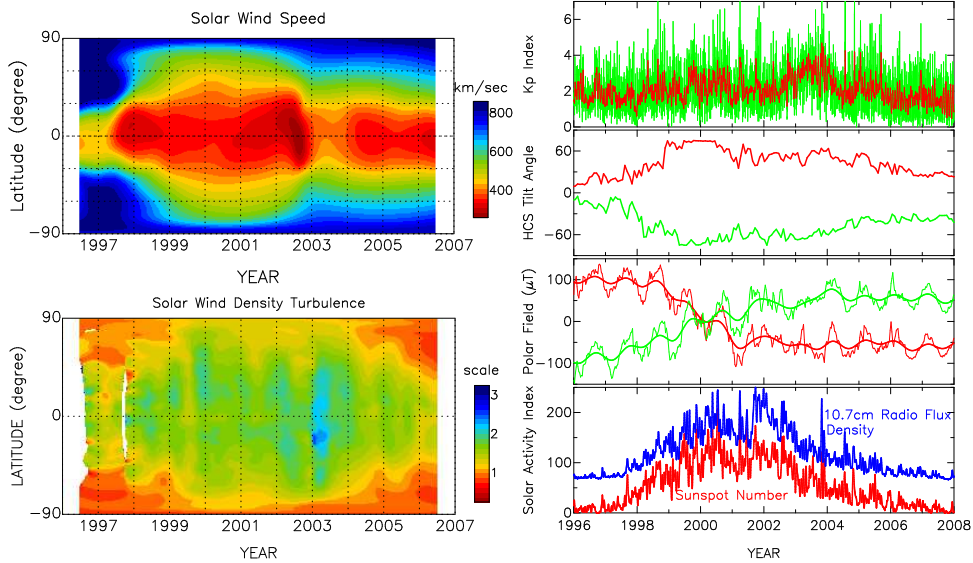}
\includegraphics[width=6.0cm]{figure_1b.ps}
\caption{(a) Latitudinal distributions of solar wind speed 
({\it left, top}) and density turbulence ({\it left, bottom}). 
The contour levels drawn on the speed gray-scale image are 350, 
450, 550, and 650 {\kmps}.
(b) The stack plot ({\it right}) shows geo-magnetic disturbance 
index, Kp, intensity of polar magnetic field, tilt angle of 
heliospheric current sheet (HCS), and solar activities (i.e., 
sunspot number and solar radio flux density at 10.7 cm).
} 
\end{center}
\end{figure}

It is evident in the `{\it latitude-year}' speed plot that during minimum
of the solar cycle, polar regions are dominated by high speed streams 
($\sim$600--800 {\kmps}) from open-field coronal holes and low and 
variable flow speeds ($\leq$500 {\kmps}) are observed at the low- and 
mid-latitude regions of the complex/closed field corona. But, there 
are marked differences in latitudinal extents of low- and high-speed 
flow regions between the current and previous minimum phases.
For example during 1996--97, the low-speed flow is confined to 
$\sim$$\pm30^\circ$ of the equatorial belt; whereas at the current 
minimum, it extends to a latitude range of $\sim$$\pm50^\circ$. These
low-speed wind widths also correlate with the latitudinal warping 
(tilt angle) of heliospheric current sheet (HCS), respectively, a 
small amplitude, $\sim\pm$15$^\circ$, at the previous minimum and a 
moderate amplitude, $\sim$$\pm$30$^\circ$, at the current minimum 
(Figure 1b). These results suggest (i) a near-dipole magnetic field
of the Sun for the previous minimum, around the year 1997 and (ii) 
a never-approached dipole-field geometry during the current minimum. 
It is in good agreement with the result from other independent IPS
speed measurements obtained from the Solar-Terrestrial Environment
Laboratory (e.g., Tokumaru et al. 2009). Thus, the HCS tilt of the 
corona of the current phase tends to resemble a condition similar 
to that of moderate activity, but without activity!

These changes in the latitudinal extents of low-speed wind have also 
influenced the high-speed flows from the polar regions. In the current 
minimum phase, high-speed regions at the poles have remarkably shrunk 
towards the poles (Figure 1a, {\it top}). Moreover, the speed of high-latitude 
($>$50$^\circ$) wind is considerably less for the current minimum than 
that of the previous minimum. These findings nicely correlate with the 
polar field strength, which is $\sim$40--50\% weaker for the current 
minimum phase. The magnetic pressure associated with the polar coronal 
holes seems to determine the acceleration of the high-speed wind. The 
weak field may be due to the fact that the polar field has not fully 
developed after the field reversal around the year 2000 (Figure 1b). 

%The large-scale features observed in the speed plot also correlate 
%with the structures seen in the density turbulence image. In Figure 1a
%({\it left, bottom}), 

\subsection{Density-Turbulence Structures}

The drifting of density structures from high to low latitudes, seen 
in Figure 1a ({\it bottom}), is caused by the slow and gradual 
movement of concentrated magnetic-field regions of the corona. It is 
likely due to the migration of small/medium-size coronal holes from 
polar to low latitude regions and the high-speed wind from these coronal 
holes interacting with the low-speed wind, causing compression in front 
of the high-speed stream.  The latitudinal spread of density patterns 
is also consistent with the HCS tilt angle (Figure 1b), which is 
maximum at the time of polarity reversal of the cycle and during when a
large number of coronal mass ejections (CMEs) dominate the heliosphere 
(e.g., Yashiro et al. 2004). 

\begin{figure}
\begin{center}
\includegraphics[width=6.7cm,angle=-90]{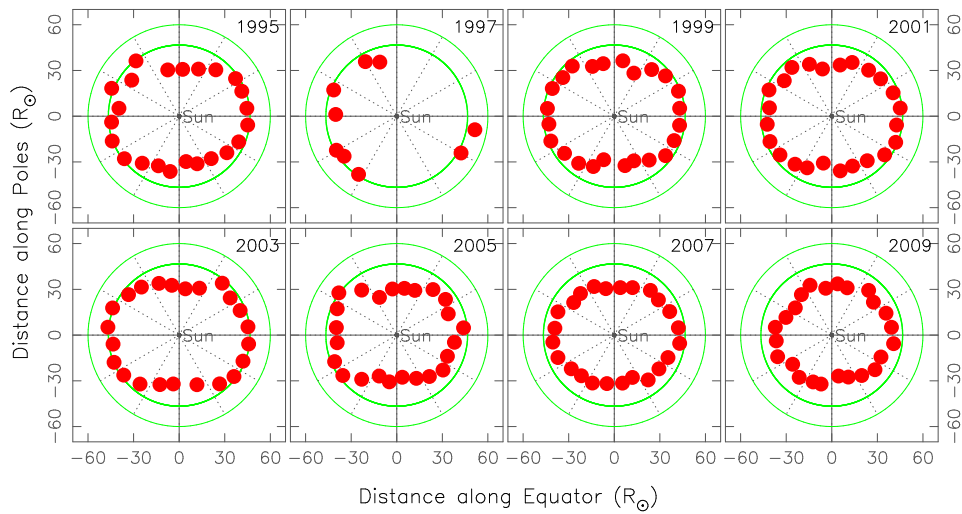} 
\caption{Shape of contours of constant density turbulence in the solar wind.}
\end{center}
\end{figure}

The density plot is also consistent with the solar wind disappearance 
period, around mid 1999 (low level of density turbulence) and 
co-rotating interaction regions (CIRs) dominated heliosphere in
the first half of 2003 (intense density turbulence along the latitudinal 
direction). It may be noted that during late October and early November 
2003, a number of CME events prevailed the interplanetary medium. However,
the effect of a CME in the Sun-Earth distance is rather limited to
2--4 days after its onset and weakens with solar distance. Whereas, 
in the case of CIR events, the influence of each event is seen for 
several days, with a systematic increase in density turbulence with 
radial distance. Moreover, they show a latitudinal pattern. For CIR 
events during 2003, the latitudinal distribution and the radial 
evolution of enhancement of turbulence have been observed (Manoharan 
2008). The CIR-dominated period, in the first half of 2003, is in 
agreement with the moderate-to-severe storms observed at the Earth 
(Figure 1b, refer to Kp index plot). As observed in the speed plot, 
the latitudinal extents of low-turbulence regions at the poles also 
show remarkable changes between the current and previous minimum phases. 
The average level of turbulence at the current minimum seems to be 
considerably lower than that of the previous cycle.

\subsection{Scattering Diameter of the Corona}

Figure 2 displays the 
constant ${\delta N_e(R)}$ plots, at different phases of the solar 
cycle. The plot for the year 1997 has been limited by weak-scattering 
observations and the last plot includes data up to May 2009.
In general, a given level of turbulence is observed closer to the Sun 
at the poles than at the equator. However, depending on the phase of
the solar cycle, the diameter of the contour can vary along the poles,
but, remains nearly the same along the equator (e.g., Manoharan 1993).

The important point to note in this analysis is that after the year 
2003, the overall diameter of ${\delta N_e(R)}$ contour has gradually
decreased with respect to the Sun's center. In other words, the same 
level of turbulence seems to move close to the Sun. Thus, the radial 
dependence of turbulence ($C^{2}_{N_e}(R)$$\sim$$[\delta N_e(R)]^2$, 
which typically varies as $R^{-4}$) suggests that the scattering 
diameter of the corona has gradually shrunk towards the Sun. In other
words, the scattering power ($C^{2}_{N_e}(R)$) has remained nearly same 
at all latitudes between 1989 and 2003, but, decreased $\sim$50\% 
around middle of 2009 at low-latitudes. 

\section{Discussion and Conclusions}

The present large-scale 3-D features of solar wind 
speed and density turbulence are remarkably different from that 
of the previous cycle. In the current minimum phase, the extent 
of low-speed region along the equatorial belt is considerably 
wider than that of the previous cycle; whereas the high-speed 
regions have shrunk towards the poles in contrast to low-latitude 
extent of the previous cycle. The other important result of this 
study is that after the year 2003, the overall scattering diameter 
of the corona has gradually decreased with respect to the Sun's 
center. These results are consistent with the ecliptic and off-ecliptic 
studies (e.g., Lee et al. 2009; McComas et al. 2008; Smith \& Balogh 
2008; Tokumaru et al. 2009).

The weak fields observed at the poles, as well as corresponding solar 
wind speed and density turbulence for the current low activity, are 
possibly caused by the changes in the movement of large-scale fields, 
as the reversal of polarity progresses. It is linked to the rate of 
poleward and equatorward meridional flows, which transport the 
unbalanced magnetic flux  (e.g., Sheeley 2008). Moreover, the 
flux-transport dynamo has predicted weak polar fields and a long solar 
cycle (e.g., Choudhuri et al. 2007). 

\begin{acknowledgments}
I thank all the members of the Radio Astronomy Centre for making 
the Ooty Radio Telescope available for IPS observations. I also 
thank National Space Science Data Center for OMNI data and the
Wilcox Solar Observatory for the magnetic-field data. This work 
is partially supported by the CAWSES-India Program, which is 
sponsored by ISRO.
\end{acknowledgments}


\begin{thebibliography}{}

\bibitem[]{} Choudhuri, A.R., Chatterjee, P., \& Jiang, J. 2007, 
             Phy. Rev. Lett., 98, 131103

%\bibitem[]{} Hewish, A., Scott, P.F., \& Wills, D. 1964, Nature, 203, 1214

\bibitem[]{} Lee, C.O, et al. 2009, Solar Physics, 256, 345 

\bibitem[]{} Manoharan, P.K. 1993, Solar Physics, 148, 153

\bibitem[]{} Manoharan, P.K., et al. 2000, ApJ, 530, 1061 

\bibitem[]{} Manoharan, P.K. 2006, Solar Physics, 235, 345

\bibitem[]{} Manoharan, P.K. 2008, in B.N. Dwivedi \& U. Narain (eds.), 
             {\it Physics of the Sun and its Atmosphere}, (World Scientific, 
             Singapore), p235-266

\bibitem[]{} McComas, D.J., et al.  2008, Geophys. Res. Lett.,  35, 18103 

\bibitem[]{} Sheeley, Jr., N.R. 2008, ApJ, 680, 1553 

\bibitem[]{} Smith, E.J. \& Balogh, A. 2008, Geophys. Res. Lett.,  35, L22103

\bibitem[]{} Swarup, G., et al. 1971, Nature Phys. Sci., 230, 185

\bibitem[]{} Tokumaru, M., et al. 2009, Geophys. Res. Lett., 36, L091001

\bibitem[]{} Yashiro, S., et al.  2004, JGR, 109, 7105


\end{thebibliography}
\end{document}